\documentclass[letter,twocolumn]{jpsj3}
\usepackage{txfonts}
\usepackage{braket}
\usepackage{color}

\usepackage{color}

\title{Tsunami Solitons Emerging from Superconducting Gap}

\author{Daisuke A. Takahashi\thanks{daisuke.takahashi@keio.jp}}
\inst{Research and Education Center for Natural Sciences, Keio University, \\ Hiyoshi 4-1-1, Yokohama, Kanagawa 223-8521, Japan} 

\abst{We propose a classical integrable system exhibiting tsunami-like solitons with a rocky-desert-like disordered stationary background. 
One of the Lax operators describing this system is interpretable as a Bogoliubov--de Gennes Hamiltonian in parity-mixed superconductors. The family of integrable equations is generated from this seed operator using Krichever's method, whose pure $s$-wave limit includes the coupled Schr\"odinger--Boussinesq hierarchy applied to plasma physics.
A linearly unstable finite background with a superconducting gap supports the tsunami-soliton solution, where the propagation of the step structure turns back at a certain moment, accompanied with the oscillation on the opposite side.
In addition, the equation allows inhomogeneous stationary solutions with an arbitrary number of bumps at arbitrary positions, which we term \textit{the Korteweg--de Vries (KdV) rocks}. In the Zakharov--Shabat scheme, the tsunami solitons are created from the Bogoliubov quasiparticles in the energy gap and the KdV rocks from normal electrons/holes. The unexpected large space of stationary solutions originates from the non-coprime Lax pair and the multivalued Baker--Akhiezer functions on the Riemann surface, formulated in terms of higher-rank holomorphic bundles by Krichever and Novikov. Furthermore, the concept of \textit{isodispersive phases} is introduced to characterize quasiperiodic multi-tsunami backgrounds and consider their classification.}

\begin{document}
\maketitle


The study of rogue waves has introduced a new interdisciplinary trend between classical integrable systems and nonlinear physics \cite{PhysRevLett.111.054101,10.3389/fphy.2020.612318,PhysRevE.85.026607}. As a prime example, the attractive nonlinear Schr\"odinger (NLS) equation has three kinds of fundamental solitons: spatially localized and temporally oscillating Kuznetsov--Kawata--Inoue--Ma soliton \cite{osti_5030825,doi:10.1143/JPSJ.44.1722,https://doi.org/10.1002/sapm197960143}, temporally localized and spatially periodic Akhmediev breather (AB) \cite{Akhmediev1986,Dudley:09}, and spatiotemporally localized Peregrine soliton \cite{Peregrine_1983}, the most famous rogue wave. 
Several integrable systems with modulational instability also exhibit similar solutions \cite{Ohta_2013,Pelinovsky2020,doi:10.1143/JPSJ.81.084001}.\\
\indent The statistical properties of integrable turbulence have also been investigated from the perspective of rogue wave formation \cite{Agafontsev_2015}. Other notable wave phenomena studied in integrable systems include the dispersive shock waves based on Whitham's method \cite{PhysRevA.74.023623}. Under these circumstances, the question arises as to whether it is possible to describe a wider range of hydrodynamic instability phenomena within the framework of the exact analysis of classical integrable systems. 
This paper proposes a classical integrable equation possessing tsunami-like soliton solutions propagating on bumpy and disordered backgrounds, emerging, surprisingly, as a byproduct of the study of superconductivity. 

\begin{figure}[tb]
		\begin{center}
		\includegraphics[width=.4\textwidth]{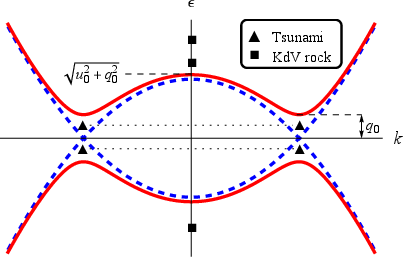} 
		\caption{\label{fig:LSCd}(Color online) Dispersion relation $\epsilon(k)$ for plane-wave eigenfunctions $\hat{L}\phi=\epsilon\phi,\ \phi \propto e^{ikx} $, with a uniform background $(u,v,q,\xi)=(u_0,0,q_0,0) $. The red solid and blue dashed lines represent the relations with ($q_0\ne0$) and without ($q_0=0$) the $s$-wave superconducting gap, respectively. The triangle and square markers show exponentially divergent seed solutions used to construct soliton solutions.}
		\end{center}
	\end{figure}

\indent Our system is described by the Lax pair
\begin{align}
	\!i\hat{L}_t&=[\hat{L},\hat{M}], \label{eq:lax} \\
	\!\hat{L}&=-\partial_x^2\sigma_3\!+\left\{\partial_x,\xi\sigma_+\!-\eta\sigma_-\right\} \!-u\sigma_3+v\boldsymbol{1}_2+q\sigma_+\!+r\sigma_-, \label{eq:laxL} \\
	\!\hat{M}&=(-\partial_x^2-u+2\xi\eta)\boldsymbol{1}_2+v\sigma_3+2(\xi_x\sigma_++\eta_x\sigma_-), \label{eq:laxM}
\end{align}
where $ \boldsymbol{1}_2 $ and $ \sigma_{1,2,3} $ are the $2\times 2$ identity and Pauli matrices, respectively, $ \sigma_\pm=\frac{\sigma_1\pm i \sigma_2}{2} $, and $\{X,Y\}$ is the anti-commutator. Here, we restrict ourselves to the case $u=u^*,\ v=v^*,\ r=q^*,$ and $\eta=\xi^*$, where $\hat{L}$ and $ \hat{M} $ become self-adjoint. The resultant equations are given by
	\begin{align}
		iu_t&=2(q^*\xi_x-q\xi^*_x),\quad iv_t=2(\xi\xi_x^*-\xi^*\xi_x)_x, \label{eq:hju} \\
		i\xi_t&=-2v\xi+q_x, \label{eq:hjxi} \\
		iq_t&=-2vq-2\xi u_x-4\xi_x u+4\xi(|\xi|^2)_x-\xi_{xxx}. \label{eq:hjq}
	\end{align}
	The physical/mathematical backgrounds that yielded Eqs.~(\ref{eq:lax})-(\ref{eq:hjq}) are described below.\\
	\indent First, $\hat{L}$ [Eq.~(\ref{eq:laxL})] can be viewed as the Bogoliubov--de Gennes (BdG) Hamiltonian \cite{Bogoliubov1958,DeGennes:1999} in parity-mixed superconductors (SCs), and the physical interpretations of the coefficient functions by the Hartree--Fock (HF) mean fields and the Cooper pairs (gap functions) are summarized in Table~\ref{ta:dict}. The mean-field theory with omitted HF fields is formulated in Ref.~\citen{Takahashi:2015nda}. The parity-mixed SCs appear in noncentrosymmetric materials \cite{doi:10.7566/JPSJ.83.013703} and the surface of topological SCs \cite{PhysRevB.90.184516}. Figure~\ref{fig:LSCd} shows a schematic of the dispersion relation with/without an $s$-wave gap $q=q_0$.\\ 
	\indent When the chemical potential $\mu$ is large, Andreev's dispersion linearization \cite{Andreev1964} around the Fermi points (see Fig. 1 of Ref.~\citen{takahashinittaJLTP}) works well. Physical systems equivalent to this one appear in diverse fields, including the Peierls problem in conducting polymers \cite{PhysRevB.21.2388,BrazovskiiGordyuninKirova,BrazovskiiKirova,BrazovskiiKirovaMatveenko,Mertsching:1981,doi:10.1143/JPSJ.52.2478,doi:10.1143/JPSJ.52.3495,BRAZOVSKII198240,DzyaloshinskiiKricheverChronek} and the Gross--Neveu models \cite{Dashen:1975xh,Shei:1976mn,Feinberg:2002nq}, and the reflectionless/finite-gap potentials in the NLS hierarchy have been applied to determine the exact self-consistent soliton dynamics and the phase diagram of the Larkin--Ovchinnikov (LO) and Fulde--Ferrell (FF) soliton lattices \cite{Machida:1984zz,Basar:2008im,Basar:2008ki,Basar:2009fg,Correa:2009xa,Takahashi:2012aw,PhysRevLett.110.131601,PhysRevLett.111.121602,PhysRevD.89.025008,PhysRevD.93.085024}.\\ 
	\indent On the other hand, the treatment \textit{without} dispersion linearization has become important in (i) the BCS-BEC crossover \cite{doi:10.7566/JPSJ.88.023601,doi:10.1143/JPSJ.76.104006}, where the BEC side corresponds to a small $\mu$, (ii) eliminating cutoff-dependence, and (iii) comparing with the quantum many-body counterpart described by the Gaudin--Yang model \cite{doi:10.1143/JPSJ.65.1227,MTakahashibook,RevModPhys.85.1633,PhysRevB.85.085414}. In particular, (iii) will play a key role in the construction of the fermionic many-body quantum solitons, whose bosonic analog has been studied \cite{doi:10.1143/JPSJ.53.1933,doi:10.1143/JPSJ.54.1710,doi:10.1143/JPSJ.55.76,PhysRevLett.108.110401,Sato_2016,PhysRevResearch.4.L032047}. The quadratic dispersion also contributes to the spectral functions and lifetime of quasiparticles in Tomonaga--Luttinger liquids \cite{doi:10.1126/science.1165403,PhysRevLett.111.256401}. \\
	\indent Next, we explain how to determine $\hat{M}$ in Eq.~(\ref{eq:laxM}). According to Krichever \cite{Krichever1977}, we can find a family of differential operators $ \hat{M}_{n,\pm},\ n=0,1,2,\dots $ commuting with $\hat{L}$, and obtain the sequence of ordinary differential equations determined by $[\hat{L},\hat{M}]=0$, called the Novikov equation. $\hat{M}_{n,\pm}$ is defined by its highest-order term $ \hat{M}_{n,+}=(-\mathrm{i}\partial_x)^n\boldsymbol{1}_2+\dotsb $ and $  \hat{M}_{n,-}=(-\mathrm{i}\partial_x)^n \sigma_3+\dotsb $. If commutators are proportional to the time derivative $\hat{L}_t$, we obtain the hierarchy of classical integrable systems.
	 Here, we use $\hat{M}=\hat{M}_{2,+}$ in Eq.~(\ref{eq:laxM}), because it provides the lowest-order equation supporting the stationary background with a superconducting gap shown in Fig.~\ref{fig:LSCd}. The \textit{Mathematica} code for generating Novikov equations is available \cite{suppl,takahashi_2025_17070336}, including expressions up to $\hat{M}_{5,\pm}$.\\
	\indent When $v=\xi=0$, the above hierarchy reduces to the one including the coupled Schr\"odinger--Boussinesq equation \cite{MAKHANKOV197442,PhysRevLett.33.148,doi:10.1143/JPSJ.81.084001,10.1143/PTP.62.370,doi:10.1143/JPSJ.57.679,MAKHANKOV19781,Krichever1994}, which has been applied to plasma physics. 
	Therefore, while the time evolution governed by $\hat{M}$ is different from those in condensed-matter systems mentioned above, whose time evolution is based on the self-consistent determination of potentials and eigenfunctions \cite{PhysRevLett.106.185301,doi:10.7566/JPSJ.88.023601,PhysRevLett.111.121602,PhysRevD.89.025008,Takahashi:2015nda,PhysRevD.93.085024} or effective field theories \cite{PhysRevD.92.034003}, the present equations (\ref{eq:hju})-(\ref{eq:hjq}) are expected to be derived by applying the reductive perturbation method \cite{MAKHANKOV197442,PhysRevLett.33.148,doi:10.1143/JPSJ.45.298,doi:10.1143/JPSJ.52.1982,doi:10.1143/JPSJ.58.4416} to multicomponent plasmas or fluids. 

\begin{table}[tb]
	\begin{center}
	\caption{\label{ta:dict} Physical interpretation of the potentials in $ \hat{L} $, describing the parity-mixed SCs. Here, $ \hat{\psi}_s,\ s=\uparrow,\downarrow $ represents fermionic field operators with up and down spin, $ g $ and $ g' $ are $s$- and $p$-wave coupling constants, respectively, and the bracket indicates the expectation value for a certain quantum many-body state.} 
	{\small
	\begin{tabular}{|c|c|c|}
	 & \multicolumn{2}{c|}{Interpretation in SCs} \\
	\hline
	$u$ & $\mu+\tfrac{1}{2}g\braket{\hat{\psi}_\uparrow^\dagger\hat{\psi}_\uparrow+\hat{\psi}_\downarrow^\dagger\hat{\psi}_\downarrow} $ & chemical potential + HF mean field \\
	$v$ & $-h+\tfrac{1}{2}g\braket{\hat{\psi}_\uparrow^\dagger\hat{\psi}_\uparrow-\hat{\psi}_\downarrow^\dagger\hat{\psi}_\downarrow} $ & magnetic field + HF mean field \\
	$q$ & $g\braket{\hat{\psi}_\downarrow\hat{\psi}_\uparrow}$ & $s$-wave Cooper pair \\
	$\xi$ & $g'\braket{\hat{\psi}_\uparrow\partial_x \hat{\psi}_\downarrow+\hat{\psi}_\downarrow\partial_x \hat{\psi}_\uparrow}$ & $p$-wave Cooper pair \\
	\hline
	\end{tabular}
	}
	\end{center}
\end{table}

\begin{figure}[tb]
	\begin{center}
	\begin{tabular}{ll}
	{\footnotesize (a)} & {\footnotesize (b)} \\[-.2em]
	\includegraphics[width=.225\textwidth]{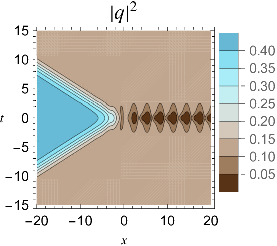}
	&
	\includegraphics[width=.225\textwidth]{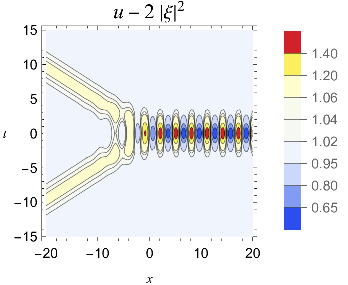} \\
	{\footnotesize (c)} & {\footnotesize (d)} \\[-.25em]
	\hspace{.3em} \includegraphics[width=.2\textwidth]{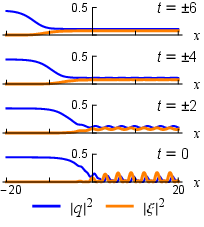}
	&
	\hspace{.3em} \includegraphics[width=.2\textwidth]{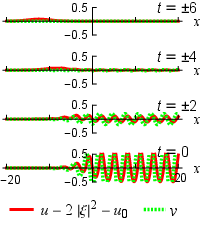}
	\end{tabular}
	\caption{\label{fig:onesol} (Color online) One-tsunami solution. The background parameters are $(u_0,q_0)=(1,\frac{2}{3})$, and the soliton parameters are $(\epsilon_1;x_1,t_1,\varphi_1)=(\frac{1}{3};0,0,0)$. (a) and (b) are contour plots for $|q|^2$ and $u-2|\xi|^2$. (c) and (d) show snapshots at special $t$ for $|q|^2,|\xi|^2, u-2|\xi|^2-u_0,$ and $v$. The gif animation for (c) and (d) is available \cite{suppl,takahashi_2025_16936981}.}
	\end{center}
\end{figure}

\indent Now, let us discuss the construction of the concrete multi-soliton solutions using the Zakharov--Shabat (ZS) scheme \cite{ZS1974}. We extract the minimal formulas for the higher-order ZS scheme\cite{TakahashiTsuda2020}: Let $ w_i,\ i=1,\dots, n $ be an eigenfunction of the Lax pair $ (\hat{L}-\epsilon_i)w_i= (\hat{M}+\mathrm{i}\partial_t)w_i=0 $ with asymptotic behavior $ \Vert w_i\Vert \to 0 $ (resp. $\infty$) at $ x\to -\infty $ (resp. $+\infty$), which are called the seed solutions. Writing their array as $ W=(w_1,w_2,\dots,w_n) $, we introduce an $n\times n$ Gram matrix $ G(x)=\int_{-\infty}^x dx W(x)^\dagger W(x) $, and define $K(x,y)=-W(x)[\boldsymbol{1}_n+G(x)]^{-1}W(y)^\dagger $.  Then, the new solution with $n$ solitons added from the known one is given by
	\begin{align}
		\xi^{\text{new}}&=\xi+K_{12}, \label{eq:dressingxi} \\
 		q^{\text{new}}&=q-2K_{12}K_{22}+(K_x-K_y)_{12}+2\xi(K_{11}\!-\!K_{22}), \label{eq:dressingq} \\
 		u^{\text{new}}&=u-2|\xi|^2+2|\xi^{\text{new}}|^2+[\ln\det(\boldsymbol{1}_n+G)]_{xx}, \label{eq:dressingu} \\
 		v^{\text{new}}&=v+\operatorname{tr}[(K_x+K_y)\sigma_3], \label{eq:dressingv}
	\end{align}
where $ K=K(x,x),\ K_x=[\partial_xK(x,y)]_{y=x},\ K_y=[\partial_yK(x,y)]_{y=x} $, and subscripts 1 and 2 represent the matrix components.
 The formulas (\ref{eq:dressingxi})--(\ref{eq:dressingv}) might be expressed elegantly using a quasideterminant \cite{Hamanaka2022}.  \\  
\indent Let us apply the above general formula to the uniform state with an $s$-wave gap $ (u,v,q,\xi)=(u_0,0,q_0,0),\ u_0,q_0>0 $. The eigenfunction with the real eigenvalue $\epsilon$ is given by
\begin{align}
	w(x,t,\epsilon,k,\varphi)=\sqrt{2|\operatorname{Re}k|}\left( w_0(\epsilon,k)e^{kx-\omega(k) t+i\varphi}+\text{c.c.} \right), \label{eq:seedw}
\end{align}
where $w_0(\epsilon,k)=\frac{1}{\sqrt{2}} \frac{1}{[(\epsilon+q_0)^2+|\omega(k)|^2]^{1/2}}\left(\begin{smallmatrix} \epsilon+q_0+i\omega(k) \\ \epsilon+q_0-i\omega(k) \end{smallmatrix}\right) $,\ $ \omega(k)= i(k^2+u_0)$,\ 
and $k=k(\epsilon)\coloneqq[(\epsilon^2-q_0^2)^{1/2}\!-u_0]^{1/2}$. 
Then, the seed solution is generally given by $ w_i=w(x-x_i,t-t_i,\epsilon_i,k_i,\varphi_i),\ k_i=k(\epsilon_i) $, possessing four real parameters $ x_i, t_i, \varphi_i $, and $ \epsilon_i $. The soliton velocity becomes $V_i = \frac{\operatorname{Re}\omega(k_i)}{\operatorname{Re}k_i}$. Below, we determine the two types of one-soliton solutions by the choice of $ \epsilon_1 $ (Fig.~\ref{fig:LSCd}). \\
 \indent The first type is determined from the superconducting gap $|\epsilon|<q_0$ (Fig.~\ref{fig:LSCd}), which we call the \textit{tsunami} soliton, whose behavior is shown in Fig.~\ref{fig:onesol}. In this solution, we observe the propagation of the step structure, which suddenly turns back at $(x,t)=(x_1,t_1)$. At the moment of turning back, the opposite side of the soliton experiences an oscillation. \\
 \indent Taking various limits of $(x_1,t_1)$, we can obtain different solutions. If both $ x_1,t_1 $ are set to infinity, we obtain a solution without the propagation turning back. If we fix $t_1$ and take $x_1 \to -\infty$, we obtain a solution with sudden oscillation occurring at $t=t_1$, similar to the AB, but it now emerges from the self-adjoint Lax pair. We note the unpredictability of the moment of turning back from the observation data---it depends on a subtle difference in the initial condition and is difficult to detect. 
 Unlike rogue waves \cite{Agafontsev_2015}, the background with finite $u_0,q_0$ is linearly unstable against a \textit{short}-wavelength perturbation; the onset of instability appears at $k=2k_F=2\sqrt{u_0}$ with a Fermi wavenumber $ k_F $, which explains the oscillation period occurring at $t=0$ in Fig.~\ref{fig:onesol}. 
 The $2k_F$-oscillation around a local defect is called the Friedel oscillation in condensed-matter context \cite{PhysRevLett.96.016801,ONO2009469}. \\
\indent Figure \ref{fig:onesol} also recalls the soliton resonance phenomena in (2+1)-dimensional integrable systems \cite{Kodama_Yeh_2016,SuyamaNagaharaMaruno201904,doi:10.1143/JPSJ.58.4416}, where Y-shaped and more divaricate structures of line solitons are formed, using large degrees of freedom, including functional parameters \cite{Matveev:1979wm}. These structures are constructed by a linear combination of multiple seed solutions, but in the present case, the maximum number of seeds is two [Eq.~(\ref{eq:seedw})] owing to the limitation of (1+1)-dimensional systems. An oscillation profile similar to Fig.~\ref{fig:onesol} can be found in two-layer fluids \cite{doi:10.1143/JPSJ.58.4416}.
 
 In Fig. \ref{fig:onesol}, we plot $u-2|\xi|^2$ instead of $u$ itself, because it is a conserved density, and it is convenient to detect the front of the tsunami soliton. The conservation laws are derived as follows \cite{doi:10.1143/JPSJ.67.1175}. Let $\Phi$ be a $2\times 2$ matrix satisfying $(\hat{L}-\lambda)\Phi=(\hat{M}+\mathrm{i}\partial_t)\Phi=0$. Defining $\Psi=\left(\begin{smallmatrix}\Phi \\ \partial_x\Phi \end{smallmatrix}\right)$, we have a zero-curvature expression $\partial_x\Psi=U\Psi,\ \partial_t\Psi=V\Psi $ using some $\lambda$-dependent $4\times 4$ matrices $U$ and $V$. Substituting $\lambda=k^2$ and expanding the Riccati equation for $\Gamma=\Phi_x\Phi^{-1}$ by powers $(\frac{1\mp i}{k})^n$, we obtain 
 \begin{align}
 	\frac{\partial }{\partial t}(\rho_n^\text{R}\pm i \rho_n^{\text{I}})+\frac{\partial }{\partial x}(J_n^\text{R}\pm i J_n^{\text{I}})=0, \quad n=1,2,\dots, \label{eq:consv}
 \end{align}
 whose first few examples are $\rho_1^\text{R}=-v,\ \rho_1^\text{I}= u-2|\xi|^2,\ \rho_2^{\text{R}}=2(|\xi|^2)_x,\ \rho_2^{\text{I}}=-v_x,\ \dots,  $ and $J_1^R=2i(\xi^*\xi_x-\xi \xi^*_x),\ J_1^\text{I}=2i(q^*\xi-q\xi^*),\ J_2^{\text{R}}=2i(\xi^*q_x-\xi q_x^*),\ J_2^{\text{I}}=2i(\xi^*\xi_{xx}-\xi\xi^*_{xx}) ,\ \dots $ etc. For a stationary solution, $J_n^{\text{R,I}}$ are constant and appear as coefficients of the algebraic curve \cite{Krichever1977}.\\

\begin{figure*}[tb]
	\begin{center}
	\begin{tabular}{llll}
	{\footnotesize (a)} & {\footnotesize (b)} & {\footnotesize (c)} & {\footnotesize (d)} \\[-.25em]
	\includegraphics[width=.325\textwidth]{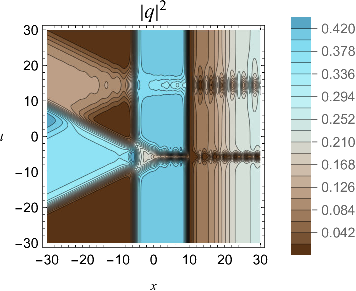}\,
	&
	\includegraphics[width=.315\textwidth]{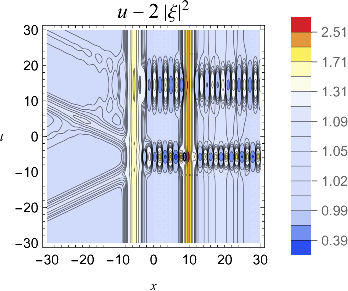}\,\,
	&
	\includegraphics[width=.13\textwidth]{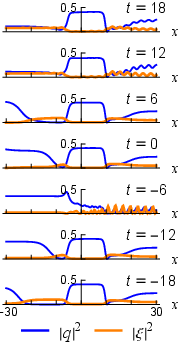}
	&
	\includegraphics[width=.13\textwidth]{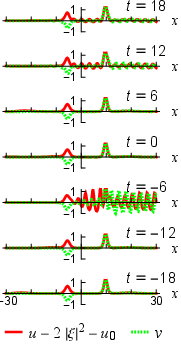}
	\end{tabular}
	\caption{\label{fig:fivesol} (Color online) Five-soliton solution with two tsunami solitons and three KdV rocks \cite{suppl,takahashi_2025_16936981}. The plotted objects are the same as those in Fig.~\ref{fig:onesol}. The parameters are $(u_0,q_0)=(1,\frac{2}{3})$ for background, $(\epsilon_1;x_1,t_1,\varphi_1)=(\frac{1}{3};0,-5,0),\ (\epsilon_2;x_2,t_2,\varphi_2)=(\frac{3}{5};-3,13,0) $ for tsunami solitons, and $(\epsilon_3;x_3,t_3,\varphi_3)=(-\frac{9}{5};-5,0,\frac{\pi}{6}),\ (\epsilon_4;x_4,t_4,\varphi_4)=(\frac{5}{2};10,0,\frac{\pi}{6}),\ (\epsilon_5;x_5,t_5,\varphi_5)=(\frac{5}{4};19,0,\frac{\pi}{6})$ for KdV rocks. The gif animation for (c) and (d) is available \cite{suppl,takahashi_2025_16936981}.}
	\end{center}
\end{figure*}

 \indent The second type of solitons arises from the bound states of normal electrons/holes (Fig.~\ref{fig:LSCd}), which has zero velocity. We call these solitons \textit{the Korteweg--de Vries (KdV) rocks}, because $\hat{L}$ with $ q=\xi=0 $ reduces to the double Schr\"odinger operator, and hence, the time evolution based on the \textit{third-order  $\hat{M}_{3,\pm}$ instead of $\hat{M}_{2,+}$} becomes the famous KdV equation. These solitons are immobile under $\hat{M}_{2,+}$, but \textit{could be} moving KdV solitons if the time evolution \textit{were} defined by $\hat{M}_{3,\pm}$. The hybrid multisoliton solution with coexisting tsunami solitons and KdV rocks is shown in Fig.~\ref{fig:fivesol}, like a flood in the desert.\\
\indent The lack of time dependence for the KdV rocks does not imply that this solution is boring---the fact that we can add an arbitrary number of stationary KdV rocks at any position is curious, because it allows the present system (\ref{eq:hju})-(\ref{eq:hjq}) to have an \textit{anomalously} large stationary-solution space. This should be compared with the typical known classical integrable systems, where general stationary solutions are given by elliptic functions and the entire solution space has at most a finite number of adjustable constants. As we will see below, this anomaly can happen because the orders of $\hat{L}$ and $\hat{M}$ [Eqs.~(\ref{eq:laxL}) and (\ref{eq:laxM})] are \textit{not} coprime. Recall that, in the KdV hierarchy, only \textit{odd-order} differential operators generate the higher-order KdV equations \cite{MiwaJimboDateEng,TanakaDate}; therefore, there is no counterpart for \textit{even-order} $\hat{M}_{n,\pm}$. 

Consider the stationary problem for Eq.~(\ref{eq:lax}), i.e., $[\hat{L},\hat{M}]=0$. By the Burchnall--Chaundy lemma \cite{https://doi.org/10.1112/plms/s2-21.1.420,a67af7a3-b821-3640-8d54-f6333ad5bc3e,Krichever1977,Krichever1978,TanakaDate,Gesztesy_Holden_2003}, the commuting differential operators satisfy a polynomial relation $P(\hat{L},\hat{M})=0$, which defines an algebraic curve (or Riemann surface). Writing the simultaneous eigenvalue problem
\begin{align}
	\hat{L}\phi=\lambda\phi,\ \hat{M}\phi=\omega\phi, \label{eq:simul}
\end{align}
the algebraic curve satisfied by (\ref{eq:laxL}) and (\ref{eq:laxM}) is 
\begin{align}
	P(\lambda,\omega)=(\lambda^2-\omega^2-a)^2+b\lambda +c\omega +d=0, \label{eq:alcrv}
\end{align}
where $ a,b,c,d $ are expressed by the rational functions of the constants $J_n^{\text{R,I}}$'s in Eq.~(\ref{eq:consv}). If the values of $a,b,c,d$ are generic, Eq.~(\ref{eq:alcrv}) represents the genus-one elliptic curve. The stationary solutions are then divided into two classes, which we call \textit{regular} and \textit{irregular} below. \\
\indent  The \textit{regular} solutions are described as follows. Assume that $J_1^\text{R}\ne0$ and set $ \alpha=J_1^\text{I}/J_1^\text{R} $ and $ \beta_3=J_3^\text{R}/J_1^\text{R} $. The stationary solution is then given by $ q=-\alpha \xi_x,\ u\pm v=\frac{1\pm\alpha}{1\mp\alpha}(\beta_3-2|\xi|^2) $, and $ (1-\alpha^2)\xi_{xx}+4(\beta_3-2|\xi|^2)\xi=0 $. The last equation is the stationary NLS equation, whose solution is given by theta functions \cite{PhysRevE.93.062224,https://doi.org/10.1111/sapm.12287}. The algebraic curve (\ref{eq:alcrv}) for regular solutions has genus $g=1$, except for the elementary limit. These results are familiar and typical. \\ 
\indent On the other hand, the \textit{irregular} solutions emerge when $J_1^\text{R}=J_1^\text{I}=0$. We have general solutions $\xi^*=c_1\xi,\ q^*=c_1 q,\ v=\frac{q_x}{2\xi} $, and $ u=\frac{c_2}{\xi^2}-\frac{q^2}{4\xi^2}+c_1\xi^2+\frac{\xi_x^2}{4\xi^2}-\frac{\xi_{xx}}{2\xi} $, 
where $ c_1,c_2 $ are integration constants. Thus, the irregular solutions have two \textit{arbitrary} real-valued (up to overall phase) functions $\xi$ and $q$, which include the multi-KdV-rock states as a particular solution. This result also suggests that the irregular solution need not generally be a reflectionless potential. With this solution, the algebraic curve (\ref{eq:alcrv}) has the squared form
\begin{align}
	P(\lambda,\omega)=(\lambda^2-\omega^2-4c_1c_2)^2=0, \label{eq:algcvirr}
\end{align}
suggesting the double-valuedness of the Baker--Akhiezer (BA) function discussed below. 

If the orders of $\hat{L}$ and $\hat{M}$ are coprime, the BA function, i.e., the simultaneous eigenfunction $\phi$ in Eq.~(\ref{eq:simul}) defined as a function on the Riemann surface, becomes a \textit{single-valued} meromorphic (excepting the essential singular point) function, whose uniqueness under a certain condition guarantees the algebro-geometric approach \cite{Krichever1977,TanakaDate,BADubrovin_1981,BelokolosBobenkoEnolskiiItsMatveev}. However, if not, it \textit{can} be (not necessarily) a \textit{multi-valued} function, and regarded as a section of the higher-rank holomorphic vector bundle \cite{Krichever1978,Krichever1979,IMKrichever_1980}. In such a case, the solution of the differential equation may have the arbitrariness of functional parameters \cite{IMKrichever_1980}.

To illustrate the above theories, let us observe that the spatially uniform state belonging to \textit{regular} solutions does not allow a time-independent KdV rock. Take the FF state $(u\pm v,q,\xi)=(p^2(1\pm\alpha)^2,-2i\alpha p\xi_0\mathrm{e}^{2ipx},\xi_0\mathrm{e}^{2ipx})$, where $ p,\xi_0,\alpha $ are real. We only discuss $\alpha=0$ for brevity and assume $p\ne0$. The eigenfunction is obtained by substituting $\phi \propto e^{i(k+p\sigma_3)x}$ into Eq.~(\ref{eq:simul}), yielding the dispersion relations $(\lambda-2kp)^2=k^4+4\xi_0^2k^2,\ (k^2+2\xi_0^2-\omega)^2=4p^2(k^2+4\xi_0^2)$. Eliminating $ k $ from these two relations, we revisit a relation of the form (\ref{eq:alcrv}), whose genus is now zero because of the elementary limit, parametrized by the uniformization parameter $s$: $ (\lambda,\omega)=(X+Y,X-Y),\ X=4p^2s(s-1),\ Y=\frac{\xi_0^2}{s}-\frac{\xi_0^4}{4p^2s^2} $. Moreover, \textit{the wavenumber $k$ is written as a rational function of the same parameter: $k=-2ps+\frac{\xi_0^2}{2ps} $,} indicating that the BA function $\phi$ is \textit{single-valued} on the Riemann surface. Indeed, $ \omega=-\lambda-4i|p|\xi_0^2\lambda^{-1/2}+O(\lambda^{-1}) $ for a large $\lambda$ has both nonzero real and imaginary parts; hence, the KdV soliton created by the ZS scheme has finite velocity, and is no longer a ``rock.'' Hence, no large arbitrariness in the stationary solution exists. 

\indent On the other hand, the $s$-wave uniform state $(u,v,q,\xi)=(u_0,0,q_0,0)$, which has been mainly considered in this study, belongs to the \textit{irregular} solution. The dispersion relations for $\phi \propto e^{ikx}$ become $ \lambda^2=(k^2-u_0)^2+q_0^2,\ \omega=k^2-u_0 $. Eliminating $ k $, we find $ \lambda^2-\omega^2-q_0^2=0 $, corresponding to Eq.~(\ref{eq:algcvirr}), which is easily parametrized as $ \lambda=\frac{q_0}{2}(s+s^{-1}),\ \omega=\frac{q_0}{2}(s-s^{-1}) $. However, \textit{the wavenumber $ k=\sqrt{\frac{q_0}{2}(s-s^{-1})+u_0} $ cannot be expressed as a rational function of the parameter $s$;} thus, the BA function $\phi$ becomes a \textit{double-valued} function on the Riemann surface owing to the square root, implying the existence of rank-2 solutions \cite{IMKrichever_1980}.

Finally, we introduce the concept of \textit{isodispersive phases} to characterize the oscillating region of the tsunami soliton (Fig.~\ref{fig:onesol}). For a given reflectionless potential of differential operator $\hat{L}$, we define the backgrounds of the left and right sides far from the potential ($x\to\pm\infty$) as \textit{isodispersive}. This term is used because both states share the same dispersion relation $ \lambda(k) $ via the Jost solution $\phi^{\text{new}}(x)=\phi(x)+\int_{-\infty}^x K(x,y)\phi(y) dy$. For spatially uniform backgrounds, isodispersive phases are typically connected by trivial gauge transformations; for example, in the integrable spinor Bose condensates with finite density \cite{doi:10.1063/1.2423222,doi:10.1143/JPSJ.75.064002}, the backgrounds before and after the soliton passes are both polar phases \cite{PhysRevLett.81.742,doi:10.1143/JPSJ.67.1822} at different angles, connected by the $U(1)\otimes SO(3)$ group \cite{10.1143/PTPS.186.455}. The same phases with various angles form the order parameter manifold, whose homotopy groups classify the topological defects \cite{Nakahara,10.1143/PTPS.186.455}. On the other hand, the tsunami soliton near the oscillation time $t\simeq 0$ (Fig.~\ref{fig:onesol}) shows a reflectionless potential such that the left side is uniform but the right side is oscillating. The multiple tsunami-soliton state (Fig.~\ref{fig:fivesol}) can support more complicated quasiperiodic backgrounds. In these cases, the set of isodispersive phases may not be compact, and the physical interpretation of the transformation group is unclear. While the uniform phases are classified by values $\rho_1^{\text{I}}=u-2|\xi|^2 $ and $ (\rho_3^{\text{I}})_{\text{uniform}}=|q|^2+\frac{u^2+v^2}{2}+2u|\xi|^2-2|\xi|^4$, characterizing all quasiperiodic isodispersive phases might require higher-order $\rho_n^{\text{R,I}}$, which remains an open problem. 

\indent In summary, we presented the tsunami-soliton and stationary KdV-rock solutions in a classical integrable equation arising from the BdG operator in parity-mixed SCs. 
The family of Novikov equations constituting a hierarchy was generated using Krichever's method. 
The tsunami solitons provide not only the turning-back dynamics but also the characterization problem of isodispersive quasiperiodic states, which might be used for new types of momentum-dependent topological defects and exotic Josephson junctions with transparent scattering properties. 
The irregular solutions, which are allowed by the non-coprime Lax pair and include the multi KdV rocks, will open up a new application of classical integrable models to physical systems with disordered backgrounds.\\ 
\indent The unified treatment of the entire hierarchy, the derivation from multicomponent plasmas and fluids by reductive perturbation, and the application to the quadratic-dispersion BdG systems toward full many-body treatment are all left as future tasks. \\
\indent The data that support the findings of this article are openly available \cite{takahashi_2025_16936981,takahashi_2025_17070336}.


%
%
%

\providecommand{\newblock}{}

{\small

}


\end{document}